# High Capacity Image Steganography Using Adjunctive Numerical Representations With Multiple Bit-Plane Decomposition Methods

James Collins, Sos Agaian

Department of Electrical and Computer Engineering
The University of Texas at San Antonio, San Antonio, Texas, USA
james.collins@utsa.edu, sos.agaian@utsa.edu

## Abstract

*LSB steganography is a one of the most widely used methods for implementing covert data channels in image file exchanges [1][2]. The low computational complexity and implementation simplicity of the algorithm are significant factors for its popularity with the primary reason being low image distortion. Many attempts have been made to increase the embedding capacity of LSB algorithms by expanding into the second or third binary layers of the image while maintaining a low probability of detection with minimal distortive effects [2][3][4]. In this paper, we introduce an advanced technique for covertly embedding data within images using redundant number system decomposition over non-standard digital bit planes. Both grayscale and bit-mapped images are equally effective as cover files. It will be shown that this unique steganography method has minimal visual distortive affects while also preserving the cover file statistics, making it less susceptible to most general steganography detection algorithms.*

## 1. Introduction

Multimedia steganography involves the means and methods by which information is embedded in a digital cover signal and communicated between two actors under the conditions that third-party observers will not be able to discern any difference between signals with embedded data and the same non-embedded original cover files[1]. One of the simplest and most popular steganographic methods involves the manipulation of the least significant bit (LSB) levels of the formatted data file [1][2]. The LSB based embedding approach is applicable to both the spatial and transform domain where least significant bits (LSB's) of digital signal values or transform coefficients can be manipulated [1]. A quick review of the common digital multimedia formats will show that there are well over three dozen main stream formats comprising both uncompressed and compressed (lossy and lossless) types that can use LSB type embedding methods successfully[2]. Operating within the trade spaces of imperceptibility, robustness, and capacity, we introduce an approach that focuses on maximizing the steganography trade space for one class of multimedia files – namely uncompressed image file formats.

In this paper, we propose a new embedding technique which alters the available number of least significant bit layers of uncompressed image files. This technique is based on the development of an entirely new redundant number system representation with subsequent remapping of the base image file to this new bit plane decomposition. Using a selective value minimization technique, data will be inserted into a number of bit planes greater than the traditional LSB levels of the first, second or third layer.  In addition, we also present the methods and algorithms necessary to demonstrate how using this novel redundant number system will increase the embedding capacity without distorting the order statistics, a necessary condition for good protection against steganalysis. The rest of the paper is organized as follows. Section 2, reviews image steganography and introduces the background on bit-plane decomposition for grayscale and bitmapped image files. Section 3 provides background on the existing works in redundant number systems and how this forms the basis for our new system, which is introduced in Section 4. Section 5 then shows a system level implementation of our approach followed by the final Section 6 covering computer simulation results.

## 2. Image Steganography

Image steganography falls under the broader classification of technical steganography what includes digital multimedia steganography.  These multimedia methods are usually listed as text, audio, image, and video embedding techniques [4].  Figure 1 shows the further breakdown of the steganography domain in the context of a larger class of cyberspace threats [5][6].

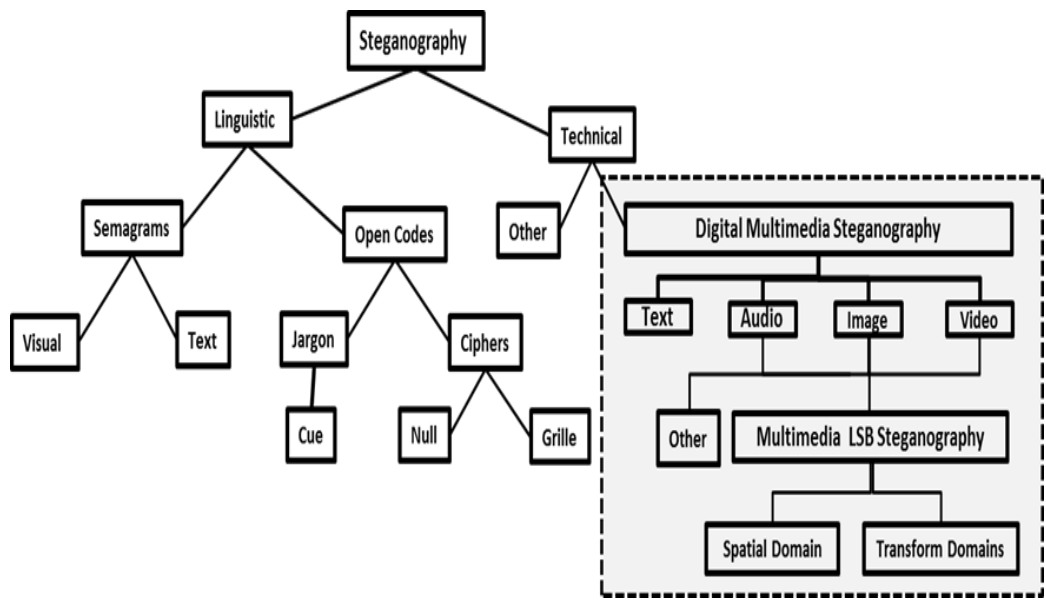

Figure 1. Expanded Taxonomy Model of Steganography Domain

Steganography aims to transmit information invisibly embedded as imperceptible alterations in common files such as images, audio, text, or video formatted as cover data [7].  Steganalysis is juxtaposed to this secret communication method with the primary objective being to unmask the presence of such hidden data.  The field of digital steganography and steganalysis continues to thrive due to the fundamental structure of digitally stored information and the covert channel bandwidth that such a structure provides [4][5].

If one considers the basic digital formats of the various multimedia files, namely audio, video, and still images, it is abundantly clear that there is considerable redundancy, variability, and fault tolerance in each of the these formats[8]. Take for example an image pixel with a range of 0 to 255. If we vary the individual sample values by several levels, the resultant change is virtually imperceptible by most observers. The human visual systems (HVS) will, in general, not notice such minor variations in a file, even if these variations are widely implemented across a given file [8][9].

The simple fact is that most of the digital multimedia formats are designed to compensate for the discrete data variability that may result from normal digital communications or processing errors [10]. It is this "flexibility" of the digital data formats that allows for steganography, watermarking, and other types of data embedding to exist for these file types. But even though these files can be used for general data embedding that would go unperceived by the HVS, any given embedding system could likely be detected by an induced statistical anomaly that is direct result of the embedding process[11][12]. Therefore, the goal of all steganography implementations is to maximize the covert channel bandwidth while minimizing the probability of third party detection [13]. The minimizations of cover distortion, both visual and statistical go hand in hand with this primary goal for steganographic techniques.

**2.1 LSB Steganography Techniques**
There are currently two major trends that are used to implement digital steganographic algorithms; those methods that work in the spatial domain, altering the desired characteristics on the file itself, and then the methods that work in the transform domain, performing a series of changes to the cover image before hiding information [2][5]]. While the algorithms that work in the transform domain are more robust, that is, more resistant to attacks, the algorithms that work in the spatial domain are simpler and faster [11]. The most popular and frequently used spatial domain steganographic method is the Least Significant Bit embedding (LSB) [11][12].

LSB embedding works by substituting message bits as the LSBs of randomly selected pixels to create an altered image called the stego-image [3]. The pixel selection is determined by a secret stego key shared by the communicating parties. Altering an LSB does not usually change the quality of image to human perception but this scheme is sensitive to a variety of image processing attacks like compression, cropping or other image translations [11][13]. Today, the majority of steganographic programs available for download from the Internet use this simple technique (e.g. Steganos II, S-Tools 4.0, Steghide 0.3, Contraband Hell Edition, Wb Stego 3.5, Encrypt Pic 1.3, StegoDos, Wnstorm, Invisible Secrets Pro. The continued popularity of the LSB embedding is most likely due to its simplicity as well as the false belief that modifications of pixel values in randomly selected pixels are undetectable because of the existence of noise present in most natural digital images [9][14].

**2.2 Bit-Plane Decomposition**
Most of these popular steganography LSB embedding tools focus on the typical 8 level bit-plane decomposition for images or amplitude level adjustments in audio formats [1][2] . These formats are all base on the standard $2^L$ layer representation, where L is the bit representations for a given binary value [1][2]. In an image for example, this typical layered decomposition is achieved using the Euclidian algorithm for the pixel values in the range of 0 to 255 [4][5]. Audio file formats are similarly ranged and the same general LSB steganography and steganalysis methodologies apply to these file types as well [15].

With the least significant bit representation, the lower level values contribute much less to overall magnitude of the specific pixel or amplitude value that the upper-most or most significant bit values convey [1][2]. For this reason, the lowest levels, say 1, 2 or even 3 are the main target for most of the 8-bit LSB data embedding tools [16][17]. We can extend the concept of LSB from 8-bit format to 24-bit color images by recognizing that the 24-bit image is merely the composite of three separate red, green and blue color components, each represented by a byte [16][17].

The widespread use of LSB embedding techniques has inspired researchers to develop a set of common steganalysis methods to detect covert data storage using these low order bit algorithms [11]. Most recently, many powerful steganalysis methods capable of detecting LSB embedding were proposed [11]. Current state-of-the-art in detection of LSB embedding is represented by the *RS analysis* and *Sample Pairs* analysis algorithms [11][18]. These analysis tools have uncovered the well-known fact that embedding near-to-maximum size messages in images using the LSB technique is quite reliably detectable by first and second order statistical analysis. However, by reducing the embedding level and effectively spreading fewer embedded bits around the cover image makes the steganalyst's task much more difficult [15][17][18]. Naturally, the steganographer is interested in developing techniques that maximize the overall embedding capacity while minimizing the probability of being detected [7].

**2.3 LSB Expansion**
Given that the current steganalysis approach is to target the structure and statistical characteristics of the first few least significant bits, alternative embedding methods are considered to avoid this type of detection. In this regard, any methods that would increase or change the number of available lower layer binary groupings would result in an expansion of the available region for embedding, improving capacity and similarly elude or evade standardized detection techniques [7].

This is the stated goal of this aspect of research and the motivation for developing alternative methods to decompose and embed information into digital multimedia files. This objective involved expanding the bit-plane decomposition beyond the standard 8-bit boundary images in an attempt to determine if alternative data embedding methods can be optimally employed. Also investigated was how these embedded schemes are affected by other than the normal bit-plane and bit-line representations. Focus was on the most basic of cases involving embedding data into uncompressed grayscale images files. This simplified, first principle approach resulted in the development of an alternative redundant number systems which was then used to formulate a unique decimal-to-binary mapping function. Our research involved in the development of this new redundant number system mapping is now reviewed.

## 3. Redundant Number Systems

As stated, many steganography methods focus on the standard 8 level bit-plane decomposition which is based on the standard $2^L$ layer representation [1][2][3]. Figure 2 shows how an image can be layer-separated to delineate the binary groupings which form an image. In reviewing these segmented images it is evident the contribution from these binary groupings increases with each successive level. That is layer 1 contributes less to the image structure than does layer 2 and much less than layers 7 or 8 [5][7]. For this reason, the lowest level 1 is the main target for most of the data embedding tools to minimize detection [11]. A goal for the steganographer is to increase the number of available lower layer binary groupings so as to increase the available regions for embedding, improving capacity and similarly eluding detection [4][5].

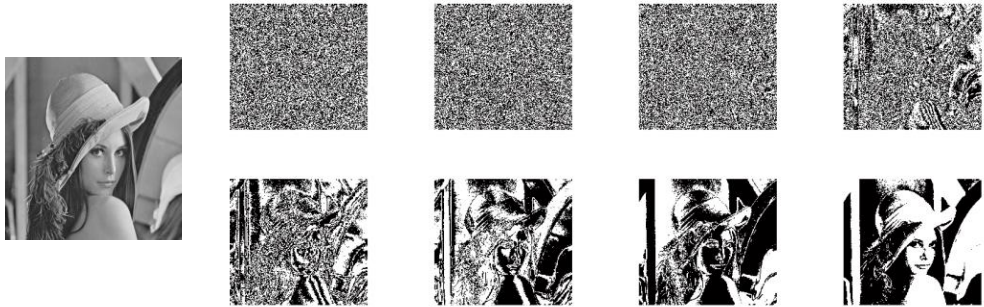

Figure 2. 8-level Bit-Plane Decomposition of Grayscale "Lena" Cover Image

One way to increase the capacity barriers associated with existing embedding systems is to find an alternate representation for the pixel value decomposition. The assuming is that in doing so we increase the available covert channel beyond the first one or two layers [19]. Our research fundamentally involves expanding the bit-plane decomposition beyond the standard 8-bit boundary to determine if alternative data embedding methods can be optimally employed. We then investigate how these embedded schemes are affected by other than the normal bit plane representations. This approach will result in the utilization of redundant number systems which will be used to form a decimal to binary mapping function. We continue by reviewing the Fibonacci redundant number system and how it is used to embed data and then introduce a new system that offers improved statistical and capacity measures.

### 3.1 Fibonacci P-Code Redundant Number System

We choose the Fibonacci series due to the deterministic qualities of the basic function. The Fibonacci *p*-code number system is described by a sequence of values generated using a fixed rule set [20]. This well-known series is a non-linear numerical sequence described by the following recursive function:

$$F_p(i) = \begin{cases} 0 & \text{if } i < 0 \\ 1 & \text{if } i = 0 \\ F_p(i-1) + F_p(i-p-1) & \text{if } i > 0 \end{cases} \quad \text{(Eq 1)}$$

Where, p is a non-negative integer designating the sequence of values particular to a given *p*. The following Table 1 shows the initial sequence of values for Fibonacci *p*-codes 0 through 3.

Table 1. Fibonacci p-codes

| p-code | Initial p-Fibonacci Numbers |
|---|---|
| 0 | 1, 2, 4, 8, 16, 32, 64, 128, 256, 512, 1024 |
| 1 | 1, 1, 2, 3, 5, 8, 13, 21, 34, 55, 89, 144, 233 |
| 2 | 1, 1, 1, 2, 3, 4, 6, 9, 13, 19, 28, 41, 60, 88, 129, 189 |
| 3 | 1, 1, 1, 1, 2, 3, 4, 5, 7, 10, 14, 19, 26, 36, 50, 69, 95, 131, 181, 250 |

A subset of these Fibonacci *p*-code values is selected from within this table with the conditions set limiting any redundant numbers in the set and bounding the set to be the minimum set of values required in a summation that is at least equal the value 255. The concatenated sets for the first three Fibonacci series are shown to be:

$$F_0(i) = \{1, 2, 4, 8, 16, 32, 64, 128\}$$
$$F_1(i) = \{1, 2, 3, 5, 8, 13, 21, 34, 55, 89, 144\} \quad \text{(Eq. 2)}$$
$$F_2(i) = \{1, 2, 3, 4, 6, 9, 13, 19, 28, 41, 60, 88\}$$

It is quickly observed that the Fibonacci p-0 series is directly related to the binary power series described by the following summation formula:

$$F_p(i) = \sum_{i=0}^{n} b_i * 2^i \quad \text{(Eq. 3)}$$

Where $b_i \in \{0,1\}$ is the coefficient for each of the bit-planes with $2^i \in \{1,2,4,8,16,32,64,128\}$. This being the case, a bijective relation can be established by pairing the *p0* Fibonacci series with an indexed binary representation to form a unique relationship between integers and the binary indexes. For example, the decimal number 13 is uniquely represented in 8-bit binary form as 1101. This bijective relation is widely known and forms the basis of all modern day digital logic arithmetic [6].

With the *p0* relation well established, we next apply similar rules to define Fibonacci-based image decomposition using the concatenated *p1* and *p2* sequences. Figure 3 shows the Fibonacci-*p1* decomposition, an 11 layer bit-plane representation. The decomposition is achieved using the Euclidean algorithm over the domain of the concatenated Fibonacci-*p1* sequence. The binary groupings for these 11 layers have a somewhat similar visual representation when compared to the standard 8-bit decomposition. Most notable is that there are now 3 lower layers that appear to be comprised of random noise-like structure. Image definition and feature vectors only begin to appear in layers 4 and 5. This aligned closely with our goal; increase the range of potential data hiding pixels [6].

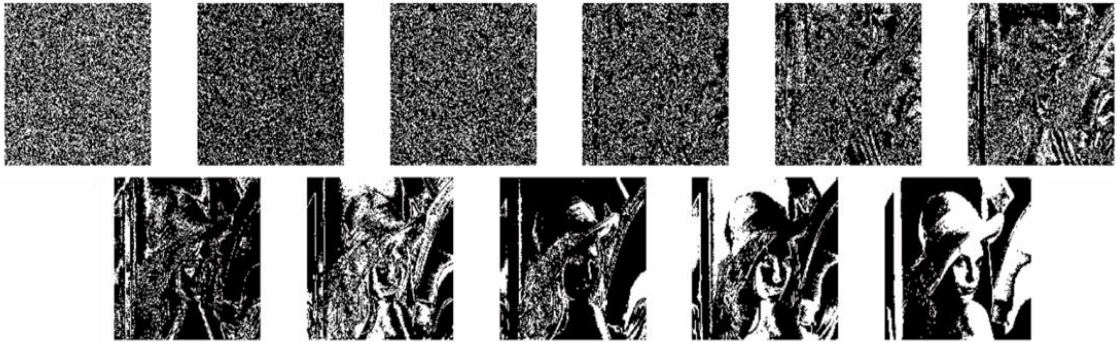

Figure 3. 11-level Bit-Plane Decomposition using Concatenated Fibonacci-*p1*

Using this same indexed binary representation method and applying it to the previously defined concatenated *p1* and *p2* Fibonacci sequences gives us a result which is a non-bijective mapping [6]. Specifically this is no longer a one-to-one relation between the decimal and indexed binary series for both Fibonacci- *p1* and *p2*. For example, the number 43 has 4 different binary index representations when mapped to the 11 bit concatenated Fibonacci-*p1* sequence:

$$43 = \{ (1,0,1,1,0,1,1,0,0,0,0), (1,0,0,0,1,1,1,0,0,0,0), (1,0,1,1,0,0,0,1,0,0,0),$$
$$(1,0,0,0,1,0,0,1,0,0,0)\} = \{a_1, a_2, a_3, a_4\}$$

We define $\{a_i\}$ as a set of binary subsets with an elemental length equal to the concatenated Fibonacci p-code set length. A decimal representation of an integer in the range of 0 to the variable n is then described by $Z = \sum_{i=0}^{n} b_i * p1_i$. Where $b_i \in \{0,1\}$ is the coefficient for each of the bit-planes and $p1_i$ is the indexed Fibonacci value for Fibonacci-*p1*. The property can be generalized for all numbers in the target domain [0, 255] by the equation:

$$A = \sum_{i=0}^{n-1} b_i * F_i \text{, Where } b_i \in \{0,1\} \text{ and } F_i \text{ is an index Fibonacci number} \quad \text{(Eq. 4)}$$

These resultant mapping gives us a surjective relationship which is expected based on the fact that Fibonacci *p1* has 11 elements and Fibonacci *p2* has 12 elements in their respective series. Naturally then, there cannot be a one-to-one mapping between *p1* and *p0* or between *p2* and *p0*, with *p0* again be our true one-to-one decimal to binary relation. This leads to the issue of how to uniquely encode and recover hidden information without exchanging any additional information, other than the encoding and decoding algorithms [6].

Several methods for solving the redundant number system uniqueness dilemma have been proposed, each with their own set of constraints [19] [20] [21]. The most often cited uniqueness scheme is based on the Zeckendorf Theorem [22]. This theorem when used for the Fibonacci sequences states that *"Each positive integer m can be represented as the sum of monotonic distinct numbers in the sequence of code numbers using no two consecutive code numbers"*. Application of the Zeckendorf theorem will in fact establish the constraints necessary to achieve translational uniqueness in a redundant number system. The aforementioned example for the number 43, translated to the Fibonacci-p1 sequence would be (1,0,0,0,1,0,0,1,0,0,0), the only sequence of the five possible $\{a_i\}$ set that satisfies the Zeckendorf constraint [22]. Clearly these results achieve a bijective relationship for our image decompositions.

Based on the progression of this Zeckendorf analysis, a valid blind embedding and decoding method for a deterministic redundant number steganography system has been achieved. However, a cursory review of this system revealed some significant implementation flaws - namely the lack of first and second order statistical preservation. These faults are of such significance in that they essentially negate one the primary purposes of steganography; that covert channel communications is difficult to detect. Experimental results which highlight these flaws will be described in the experimental results section below.

We continued then, with our basic research to find a redundant number system that would allow us to embed more information over the decomposed image layers while simultaneously reducing the chances of detection. These objectives lead us to the development of a new class of specialized number systems which we introduce as *Adjunctive Numerical Representations*.

## 4.0 Adjunctive Numerical Representation
It can be shown that the mapping function from a decimal to binary index relation for Fibonacci-*p1* is [0, 375] and is [0, 284] for our Fibonacci-*p2* set, both surjective functions. It was also notable that a one-to-one mapping exist between the domain of numbers 0 to 255 and our concatenated Fibonacci-*p0* series which then forms the basis function used in modern day digital computing. Table 1 defined the p Fibonacci values for *p-0 to p-3*. The full range of values for an image color value [0, 255] can be expressed using p-0 to p-2. The summation curves for these three series are plotted and shown in Figure 4.

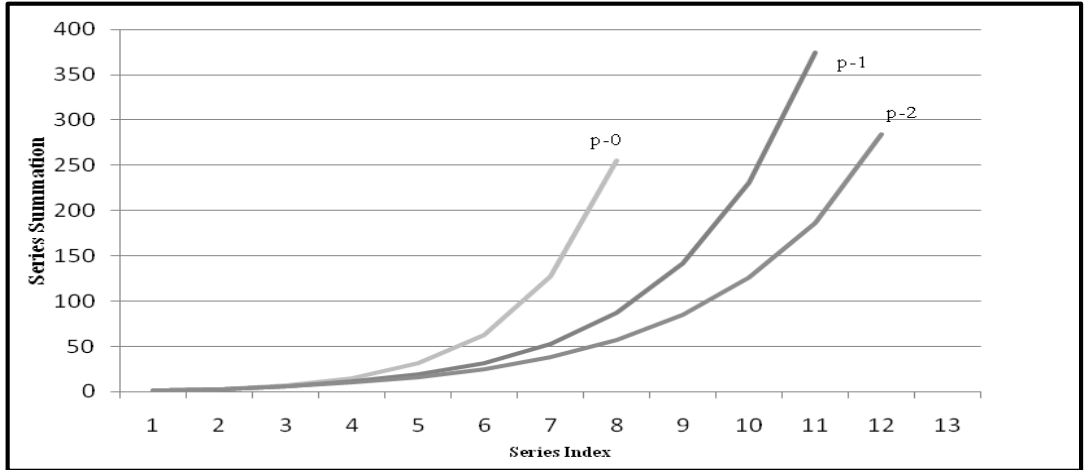
Figure 4. First Three Fibonacci *p*-code Summation Plots

From this output, it can be seen that the ideal curve is *p0* since this is the bijective relation that is necessary for proper integer domain mapping. Curves p1 and p2 peak at 375 and 274 respectively and are not the ideal bijective relational systems, even when the constraints of the Zeckendorf theorem are applied. We also saw that the p-1 and p-2 embedding system have non-ideal results due to distortions in the order statistics [19][20][21][23]. The objective is to design a redundant number system (required for multiple bit-plane decomposition) that would closely model the characteristics of the idea Fibonacci-*p0* and yet continue to preserve the critical digital cover file statistical values.

We now define those characteristics deemed necessary for our new set of indexing number sequences, heretofore referred to as *Adjunctive Numerical Representation* (ANR) sets. The following five properties define the uniqueness for $ANR_N$ set of numbers:

I. $ANR_N \stackrel{\text{def}}{=} [0, \sum_{i=1}^{n} x_i] = [0, N]$, where , $x_i < x_{i+1}$

II. $\sum_{i=1}^{n} x_i = N \ \forall \ ANR_N$, where $\{ANR_N | (x_1, x_2, x_3, \ldots x_n)\}$

III. $\forall \in [0, N], \{T: ANR_N \rightarrow F_p(i)\} \ \nexists! \ \{a_i\}$, where $F_p(i) = \sum_{i=0}^{n} b_i * 2^i)$, $b_i \in \{0,1\}, p \in Z$

IV. $\forall \in [0, N] \ in \ ANR_N, \exists \min \{T: ANR_N \rightarrow F_p(i)\} = min\{a_i\}$

V. $\forall \in [0, N] \ in \ ANR_N, \exists \max \{T: ANR_N \rightarrow F_p(i)\} = max\{a_i\}$

Property one is derived from the ideal case of Fibonacci-*p0* where the series summation results in the value 255, the upper bound of the 8-bit digital boundary. The second property ensures that all the natural numbers between 0 and N can be represented by the redundant number system $ANR_N$. The third property is a necessary condition to ensure the indexed values form a redundant number system. Specifically, this third property ensures that there will be multiple representations of a given number, thus aiding in the obfuscation process during data embedding. Finally, the last two properties are fundamental rules needed for blind system encoding and decoding, essentially stating that a distinct minimum and maximum element must exist for a distinct integer representation [19][22][23]. These five properties will then bring us close to the ideal Fibonacci-p0 case but will lack the totality of it unique properties of the binary number system. This coupled relationship is used to derive the name for our new system: Adjunctive Numerical Representation (ANR).

We identify the following sets of numbers as representative subsets for our new $ANR_{255}$ redundant system. These monotonic sets are defined to be consistent with the aforementioned number system properties. The selection of an ANR image decomposition set is based on the five properties previously defined.

$$S1 = \{1, 2, 3, 4, 6, 10, 14, 17, 23, 31, 42, 47, 55\}$$
$$S2 = \{1, 2, 3, 4, 5, 6, 7, 8, 15, 36, 43, 57, 68\}$$
$$S3 = \{1, 2, 3, 7, 11, 13, 17, 21, 23, 27, 31, 41, 58\}$$

The reader can easily verify that $\sum_{i=1}^{13} S1_i = \sum_{i=1}^{13} S2_i = \sum_{i=1}^{13} S3_i = 255$ and that the set of all integers from 1 to 255 can be represented by each of these indexed sequences. The identification of these three ANR sets shows that we have expanded the number of available image decomposition bit planes from the standard 8 to 13.

The plots for these three 13-level sets are shown below in Figure 5 alongside the previously plotted curves. Observe that all three of these $ANR_{255}$ number sets converge to 255. Note as well that the curves also closely model the ideal curve of Fibonacci-$p0$ over the initial indexing interval of approximately 1 to 9. Property three is verified by noting the number of elements, 13-levels, in each of the $ANR_{255}$ series, is greater than the 8 found in Fibonacci-$p0$. The final two properties define the uniqueness requirement for our new system. The development of these two properties is described in the following section.

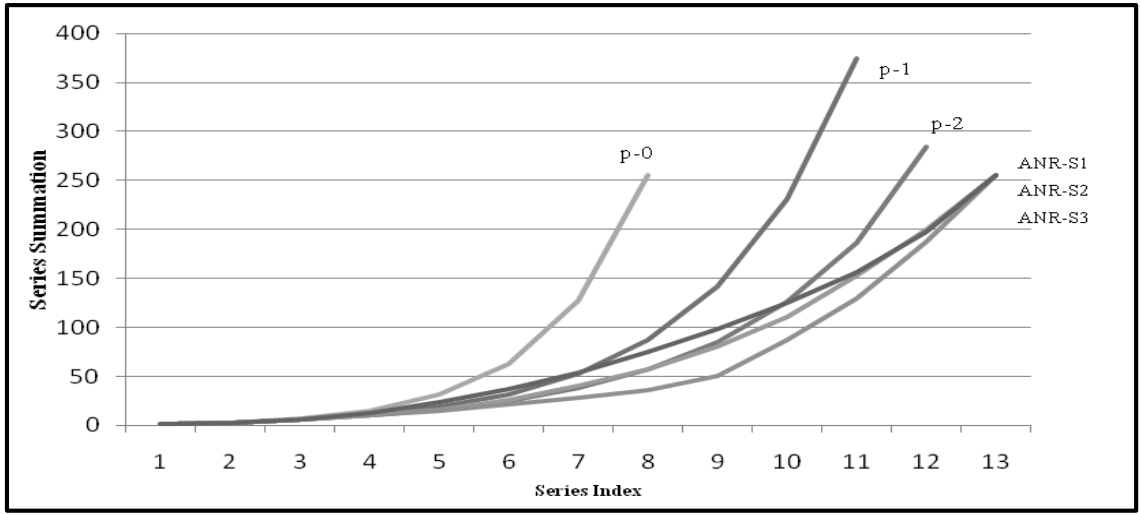

Figure 5. Fibonacci $p$-code and Representative $ANR_{255}$ Summation Plots

### 4.1 Uniqueness in $ANR_{255}$ Representations

From the above sections, we identified the need of unique numerical representation for various redundant number systems. Specifically we want to find uniqueness in both the Fibonacci and $ANR_N$ redundant number systems. The notion of a unique representation motivated us to investigate various coding techniques and develop a Min-Max bit-plane normalized representation that helps us in designing our blind embedding system [22][23]24]. For this we postulate the following:

> Theorem: *Any natural number, Z, within an interval [0, 255] can be represented and uniquely identified by a set of monotonic numbers over a concatenated Fibonacci p-code or and $ANR_N$ set with the binary sequence mapping, $a_i$ selected*

> *using the maximum representation, $max(\{a_i\})$, or minimum representation, $min(\{a_i\})$, where, $\{a_i\}$ is one of the possible decimal-to-binary mappings in the initial translation*[23][24].

The maximum and minimum representations of each $\{a_i\}$ in the redundant number system are derived from the Euclidian reduction of a given decimal number when represented as indexed binary form. Once in this form the minimum and maximum are determined by lexiconical ordering. An as illustrative example of this uniqueness theorem, consider the concatenated Fibonacci-*p1* and the ANR$_{255}$ representations of the number 43. Previously we showed that there were 5 representations of this number when mapped into the 11-bit sequence of the previously defined Fibonacci-*p1* series. The simple maximum and minimum selection functions define uniquely represented values as shown in Table 2. For the ANR$_{255}$ system the number 43 is mapped into 9 different 13 bit binary sets as shown in Table 3.

Table 2. Fibonacci-*p1* Representation and Uniqueness

| Integer value 43 in representation $\{a_i\}$ form for concatenated Fibonacci-*p1* P1={1,2,3,5,8,13,21,34,55,89,144} | $max(\{a_i\})$ | $min(\{a_i\})$ |
|---|---|---|
| $a_1 = \{1,0,1,1,0,1,1,0,0,0,0\}$<br>$a_2 = \{1,0,0,0,1,1,1,0,0,0,0\}$<br>$a_3 = \{1,0,1,1,0,0,0,1,0,0,0\}$<br>$a_4 = \{1,0,0,0,1,0,0,1,0,0,0\}$ | $a_4 = \{1,0,0,0,1,0,0,1,0,0,0\}$ | $a_1 = \{1,0,1,1,0,1,1,0,0,0,0\}$ |

Table 3. ANR$_{255}$ Representation and Uniqueness for Integer 43

| Integer 43 representation in $\{a_i\}$ form for new ANR$_{255}$ S1={1,2,3,4,6,10,14,17,23,31,42,47,55} | $max(\{a_i\})$ | $min(\{a_i\})$ |
|---|---|---|
| $a_1 = \{1,1,1,0,1,0,1,1,0,0,0,0,0\}$<br>$a_2 = \{0,1,0,1,1,0,1,1,0,0,0,0,0\}$<br>$a_3 = \{0,1,0,0,0,1,1,1,0,0,0,0,0\}$<br>$a_4 = \{1,1,0,0,0,0,0,1,1,0,0,0,0\}$<br>$a_5 = \{0,0,1,0,0,0,0,1,1,0,0,0,0\}$<br>$a_6 = \{1,1,1,0,1,0,0,0,0,1,0,0,0\}$<br>$a_7 = \{0,1,0,1,1,0,0,0,0,1,0,0,0\}$<br>$a_8 = \{0,1,0,0,0,1,0,0,0,1,0,0,0\}$<br>$a_9 = \{1,0,0,0,0,0,0,0,0,0,1,0,0\}$ | $a_9 = \{1,0,0,0,0,0,0,0,0,0,1,0,0\}$ | $a_1 = \{1,1,1,0,1,0,1,1,0,0,0,0,0\}$ |

To this point we have described some fundamental concepts of redundant number systems. We showed how a set of Fibonacci p-code redundant number systems could be developed into a multiple bit-plane encoding and decoding system by adding the Zeckendorf uniqueness constraints [22]. We then expressed without proof that these system would fail under steganalysis using 1[st] and 2[nd] order statistical attacks, the substantiating results which will be shown in the experimental analysis below. A compilation of the aforementioned results lead us to develop the new redundant system which embodied the attributes necessary for a high capacity, low statistical distortive embedding algorithm using our newly defined adjunctive numerical relationship principles. In the following two sections we describe in general terms how this new system is implemented and then present some output from computer simulations with supporting data that demonstrates the effectiveness of our new ANR embedding system.

## 5. System Implementation

In this section we present the encoding and decoding process which uses the new multiple bit-plane decomposition ANR algorithm [6][15]. Figure 6 depicts the individual steps involved in our stegonographic system. The first step in the process is to decompose the image into a fixed set of bit planes. The number of planes to be represented is determined by the number of index values for a given $ANR_{255}$ sequence. For example, one of the sequences that adheres to the properties previoulsy defined for our system was sequence S1. This was defined by the set {1,2,3,4,6,10,14,17,23,31,42,47,55} with unique 13 elements. The image then will be decomposed into 13 bit planes with the lowest order being 1 and a highest order plane of 55. The conjugate of this in normal image decomposition would be 8 bit planes with 1 as the lowest and 128 as the highest order bit plane. Figure 7 shows $ANR_{255}$ decomposition of the Lena image into the 13 layers as defined by the S1 sequence [15].

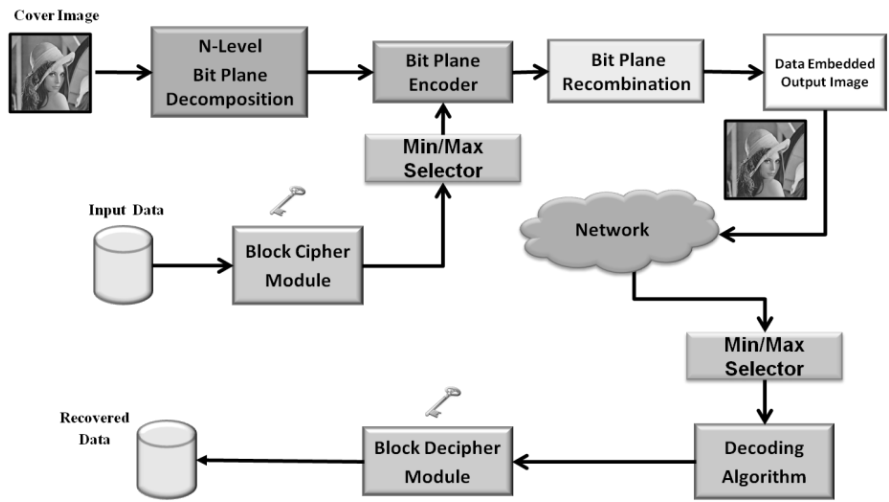

Figure 6. Decomposition Embedding and Decoding System

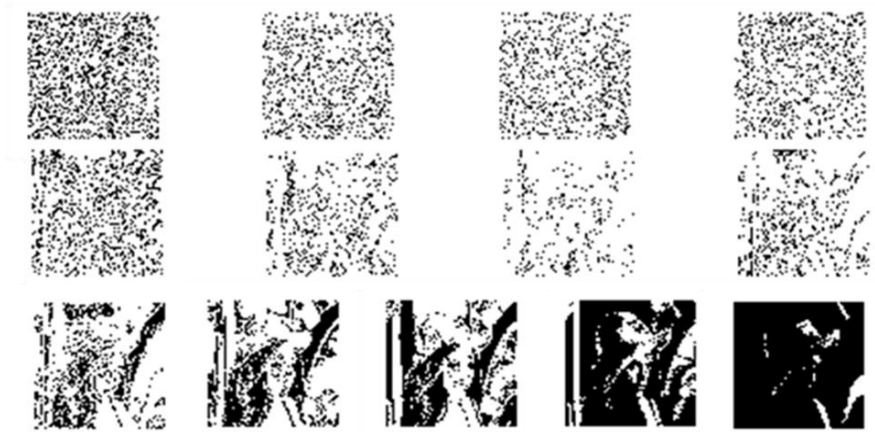

Figure 7. $ANR_{255}$ -S1 13-Level Bit-Plane Decomposition

## 5.1 Embedding and Extraction Algorithms

Once the image is properly decomposed, preprocessed encrypted data is ready to be stream embedded into the first three levels of these separated bit planes. The location of the embedded data is governed by the $ANR_{255}$ conversion of the image intensity value as described in Table 3.3 with the selection of *min* or *max* being applied to this number set. The insertion of this streaming data can be either adaptively or non-adaptively embedded; the choice of which is guided by several factors including the amount of data to be embedded and the desired security of the overall system. The selection on which bit-plane to embed is based on the vector set from which the sample data resides. Figure 8 shows the ANR embedding code with the associated vector elements. Once the data is embedded into layers one, two, and three of the 13-layer bit-plane decomposition, the cover file is converted to the standard 8 bit-plane representation. The decoding process is simply performed by reading the decimal-based image intensity values and extracting the data directly using the appropriate *min* or *max* representation of this number in $ANR_{255}$ format.

```
for ii = ih1+2:blocksize:M-ih1
  for jj = jh1+2:blocksize:N-jh1
    sum = 0;
    Neighborhood = blankImage(ii-ih1:ii+ih1, jj-jh1:jj+jh1);
    Window1_D = reshape(Neighborhood, 1, WindowLength);
    L = length(Window1_D);

    Neighborhood1 = dataImage(ii-ih1:ii+ih1, jj-jh1:jj+jh1,1);
    Window1_D1 = reshape(Neighborhood1,1,WindowLength);

    Neighborhood2 = dataImage(ii-ih1:ii+ih1, jj-jh1:jj+jh1,2);
    Window1_D2 = reshape(Neighborhood2,1,WindowLength);

    Neighborhood3 = dataImage(ii-ih1:ii+ih1, jj-jh1:jj+jh1,3);
    Window1_D3 = reshape(Neighborhood3,1,WindowLength);

  for step=1:WindowLength
    if index >= DataLength
      break;
    else
      index = index + 3;
    end
    Window1_D1(step) = bitset(Window1_D1(step),1,dataVector(index));
    Window1_D2(step) = bitset(Window1_D2(step),1,dataVector(index+1));
    Window1_D3(step) = bitset(Window1_D3(step),1,dataVector(index+2));
  end
 end
  result = double( sqrt(sum) / WindowLength);
end

    vector1 = [13 14 15 16 17 18 19 20 37 38 39 40 77 78 79 80 85 86  87 88 113 114 115 116 133
               134 135 136 197 198 199 200  205 206 207 208 221 222 223 224 229 230 231 232];

    vector2 = [54 55 56 57 98 99 100 101 102 103 104 105 150 151 152 153 166 167 168 169 174
               175 176 177 210 211 212 213 242 243 244 245 246 247 248 249];

    vector3 = [3 4 5 6 7 8 9 10 23 24 25 26 27 28 29 30 31 32 33 34 67 68 69 70 71 72 73 74 91 92
               93 94 119 120 121 122 127 128 129 130 155 156 157 158 187 188 189 190 191 192
               193 194 215 216 217 218 235 236 237 238];

    vector4 = [44 45 46 47 48 49 50 51 60 61 62 63 108 109 110 111 140 141 142 143 144 145 146
               147 160 161 162 163 180 181 182 183 252 253 254 255];

    vector5 = [0 1 11 12 21 22 35 36 42 43 52 53 58 59 65 66 75 76 81 82 83 84 89 90 96 97 106 107
               117 118 123 124 125 126 131 132 138 139 148 149 164 165 170 171 172 173 178 179
               185 186 195 196 201 202 203 204 219 220 225 226 227 228 233 234 240 241 250 251];
```

Figure 8.  $ANR_{255}$ -S1 13-Level Bit-Plane Embedding

# 6. Computer Simulation

In order to assess our new Ajunctive Numerical Representation algorithm for resistance to first and second order steganalytic attacks and measure the embedding capcitity, computer simulations were performed over a small database of 30 grayscale images. The set of files listed below in Table 4, and shown in Figure 9 comprise a representative subset of the image files used in our experimentation.

Table 4. Representative Image Set

| Representative Image Set | Size | Embedding Capacity Per Bit-Plane | Baseline Stego Sensitivity Measure |
|---|---|---|---|
| **Lena** | 512x512 | 262,144 | 0.000 |
| **Barbara** | 510x510 | 260,100 | 0.000 |
| **Baboon** | 512x512 | 262,144 | 2.307 |
| **Peppers** | 512x512 | 262,144 | 0.137 |

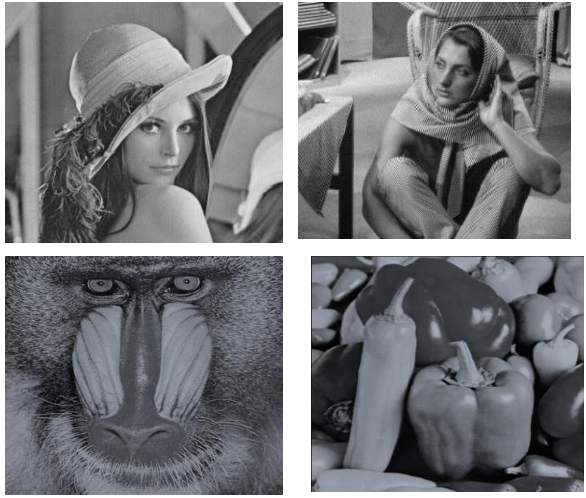

Figure 9 ANR$_{255}$ -S1 Vector 5 Embedding for 3 bit-planes

The table shows the image size, and the bit level embedding capacity for each of the bit planes. For our experments we will be using grayscale images and will limit our total embedding capacity to one full capacity LSB level. Table 4 also provide a baseline measurement of the clean images using the Stego Sensitivity Measure (SSM) algorithm [25][26].

The SSM algorithm is designed to detect evidence of localized visual artifacts as an indication of steganographic content in the digital image [25]. The methods involves the measurement and classification of pixels which are adjacent to a center focus pixel. Variations from this central data point by the surrounding pixels, when proportionally measured, are used to determine if a given pixel values has a likelihood of covert information [25][26]. If we consider an image *I* of dimension *MxN*, we may define a sub-element block, $I_b$ to be dimension $m_b$ by $n_b$. From this sub-element selection, the adjacent pixel values to the center pixel will be:

$$| P_{ij} + k - P_{i\pm 1\ i\pm 1}| \leq A \qquad \text{(Eq. 5)}$$

where A is the defined threshold of the bit-plane under analysis, and P are the pixel values.

The value k is ranged from: -A, -A+1,…0,…A-1, A [25][26][27]. The pixel comparison complexity measure is defined by the relationship:

$$\gamma(m_b, n_b) = \frac{\beta}{\beta_{max}} \quad \text{(Eq. 6)}$$

Where $m_b$ x $n_b$ are the blocks under analysis. The value β shows the count of the adjacent pixel pairs meeting the threshold and βmax is the total adjacent pairs [25]. The global value summarizes the relationships between these pixel measures and is define as the complexity measure of the image.

$$\Gamma = \frac{1}{B}\sum_{b=1}^{B} \gamma_b (m_b, n_b) \quad \text{(Eq. 7)}$$

Typically, an SSM measure falling below 5 is considered to be a clean image, a value between 5 and 10 is deemed suspicious and measured value above 10 is usually identifies a cover image with steganographic content [25][26].

We used the SSM measure for the initial unembedded image as a baseline. Once the image is processed with a set of embedding algorithms, we can correlate the resultant SSM value with the baseline in order to make an accurate judgment of the results.

In addition to the SSM, three additional image post-process parametric analysis values were collected. These include the Structural Similarity Measure, SSIM, the peak signal-to-noise ratio or PSNR, and the mean squared error or MSE [28]. The SSIM is an improved comparative analysis measure when measured against the PSNR or MSE. The method for calculating this index involves the definition of two windows on the image. The measurement between these windows of size NxN is given by:

$$SSIM(x, y) = \frac{(2u_x u_y + c_1)(2\sigma_{xy} + c_2)}{(\mu_x^2 + \mu_y^2 + c_1)(\sigma_x^2 + \sigma_y^2 + c_2)} \quad \text{(Eq.8)}$$

of which the values are represented as:

$$\mu_x = average\ of\ x \quad \mu_y = average\ of\ y$$

$$\sigma_x^2 = variance\ of\ x \quad \sigma_y^2 = variance\ of\ y \quad \text{(Eq. 9)}$$

$$\sigma_{xy} = covariance\ of\ x\ and\ y$$

$$c_1 = (k_1 L)^2,\ c_2 = (k_2 L)^2 \quad with\ k_1 = 0.02\ and\ k_2 = 0.03\ \text{as a basis}$$

Even though the SSIM is an improvement over other methods it has not achieved a full acceptance by the image processing community. For this reason, we also provide in our study, the classic PSNR and MSE values [29].

The PSNR is power ratio measurement for the absolute values between two measured signals. This measure is useful in that it can show the effects of noise or signal degradation on a systems output following a given digital signal process. We can then define the signal power between two images with the following relationship:

$$PSNR = 10 * log_{10}\left(\frac{MAX_I^2}{MSE}\right) \quad \text{(Eq. 10)}$$

In the case of our image measures, $MAX_I$ is the maximum possible pixel value of the image which for a given bit-plane is 255. Notice that the PSNR uses the calculated mean square error when calculated directly between pixel images [30]. This result, MSE, is defined as:

$$MSE = \frac{1}{MN}\sum_{i=1}^{M}\sum_{j=1}^{N}[I_1(i,j) - I_2(i,j)]^2 \quad \text{(Eq. 11)}$$

For two images of size $M$ x $N$ iterated over $i$ rows and $j$ columns.

### 6.1 ANR Direct Embedding Results
With each of the parametric measurements define, we now review the results our new ANR embedding technique as compared to other widely used LSB embedding methods. To test the effectiveness of proposed ANR embedding algorithm we compared it to a set of LSB embedding functions that target data for the 1st, 2nd and 3rd bit-planes. Data was embedded into each of the 30 database test images at 10%, 25%, 50% and 100% of the individual images as measured by the least significant bit capacity as shown previously in Table 4. The LSB embedding method used was direct encoding, essentially resulting in a +/- or bit flipping function to that matches the covert data value. Before embedding, data was randomized as a fixed stream of binary data. This ensures that a smooth probability distribution is enforced so as not to skew the original image histogram profile.

Table 5 details the results for the *Lena* image. A complete set of results for each of the four representative figures is provided in Appendix 1. As we interpret these results we begin with the initial baseline measures for the SSM shown below as 0 and 215.53 respectively. The most significant value we extract from this data set is the SSM results. The SSM, initially being 0 is increased only slightly in ANR with 100% embedding. Given that the SSM is one of the most effective steganalysis measures against LSB based embedding algorithms an ANR value of 0.22 is well within the margin of non-detection. The calculated SSM values for a majority of the LSB measures shows markedly elevated values that would be highly indicative of covert data present in the image. In general, the PSNR and SSIM values are not of any particular consequence. There is a slight drop in the PSNR for 100% embedding using 3 level LSB. However, in this instance with a SSM value of over 28, the image would be flagged regardless.

The probability of embedded data detection, *P(S)*, is zero in all cases for ANR. There are some significant peaks in this value for the LSB embedding, especially when embedding rates exceed 50% of capacity. This value, along with the positive $Chi^2$ indicate that the ANR embedding algorithms has good second order statistical characteristics [29][30]. The results shown in this table indicate the ANR embedding algorithm is an improvement over traditional direct encoded LSB embedding methods.

Table 5. ANR Embedding Results for LENA Image

| Lena SSM Base= 0 , Chi2 = 215.53 | | | | | | | | |
|---|---|---|---|---|---|---|---|---|
| 10% | Total Bits | SSM | SSIM | PSNR | Chi2 | %Diff | P(S) | % |
| 1LSB | 26215 | 11.54 | 0.9999 | 140.73 | 155.78 | -38.35 | 0.0005 | 0 |
| 2LSB | 26215 | 3.79 | 0.9997 | 131.53 | 210.59 | -2.35 | 0.0000 | 0 |
| 3LSB | 26215 | 0.67 | 0.9995 | 121.51 | 210.40 | -2.44 | 0.0000 | 0 |
| ANR | 26215 | 0.00 | 0.9997 | 132.67 | 230.44 | 6.47 | 0.0000 | 0 |
| 25% | Total Bits | SSM | SSIM | PSNR | Chi2 | %Diff | P(S) | % |
| 1LSB | 65536 | 26.06 | 0.9997 | 131.65 | 155.78 | -38.35 | 0.0421 | 4 |
| 2LSB | 65537 | 11.25 | 0.9993 | 122.44 | 183.66 | -17.35 | 0.0008 | 0 |
| 3LSB | 65536 | 4.73 | 0.9979 | 112.14 | 196.50 | -9.68 | 0.0001 | 0 |
| ANR | 65536 | 0.00 | 0.9993 | 123.59 | 244.88 | 11.99 | 0.0000 | 0 |
| 50% | Total Bits | SSM | SSIM | PSNR | Chi2 | %Diff | P(S) | % |
| 1LSB | 131072 | 47.93 | 0.9995 | 124.69 | 113.04 | -90.67 | 0.8073 | 81 |
| 2LSB | 131073 | 22.92 | 0.9986 | 115.58 | 160.58 | -34.22 | 0.0235 | 2 |
| 3LSB | 131074 | 12.32 | 0.9959 | 105.23 | 177.99 | -21.09 | 0.0019 | 0 |
| ANR | 131073 | 0.00 | 0.9988 | 116.60 | 267.74 | 19.50 | 0.0000 | 0 |
| 100% | Total Bits | SSM | SSIM | PSNR | Chi2 | %Diff | P(S) | % P(S) |
| 1LSB | 260101 | 100.00 | 0.9992 | 117.85 | 55.92 | -285.46 | 1.0000 | 100 |
| 2LSB | 262145 | 44.57 | 0.9976 | 108.62 | 114.82 | -87.71 | 0.7727 | 77 |
| 3LSB | 262144 | 28.03 | 0.9927 | 98.35 | 131.55 | -63.84 | 0.3731 | 37 |
| ANR | 262144 | 0.22 | 0.998 | 109.58 | 355.60 | 39.39 | 0.0000 | 0 |

When the histograms are considered for these results there does not appear to be any specific or dominant indicators as markers for clear steganographic presence. Shown below in Figure 10 is the original unembedded histogram for the *Lena* image. Figure 11 is the Lena image histogram with 100% embedding using the ANR method. The final figure, 12 is the same image with 100% direct embedding using LSB up to Layer-3. While some slight peak anomalies exist in all three images, detection using double blind histogram analysis of these would graph, by most estimations, would be inconclusive. Using automated statistical detection engines, the probability of embedded data detection, *P(S)*, is zero for the ANR embedded image but is 37% for the 3-level LSB embedding technique.

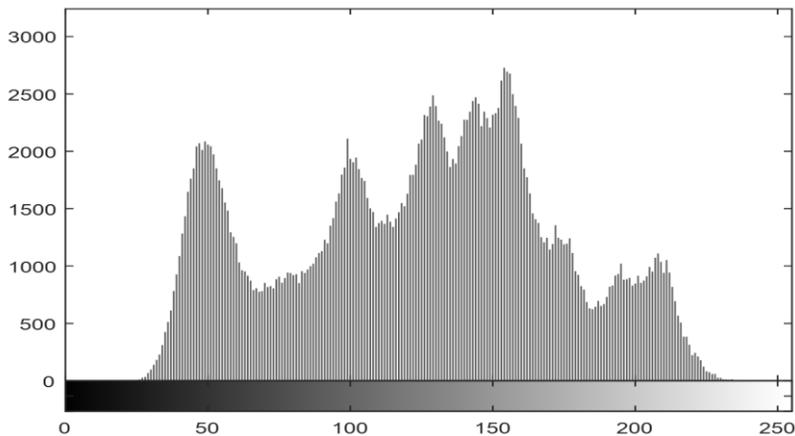

Figure 10 Original LENA Image Histogram

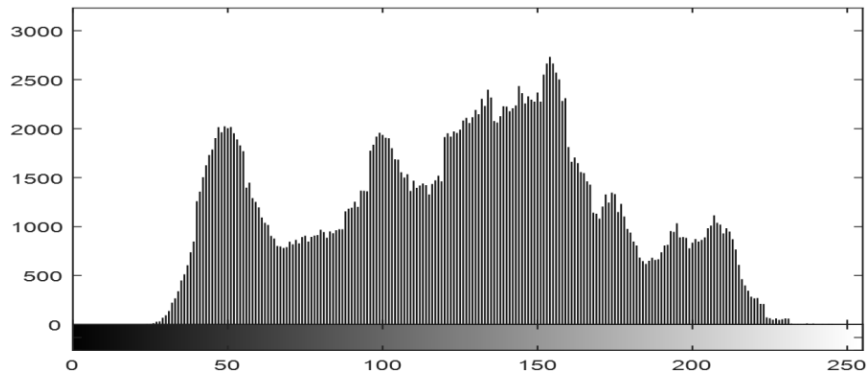

Figure 11 LENA Image Histogram for 100% LSB Direct Embedding

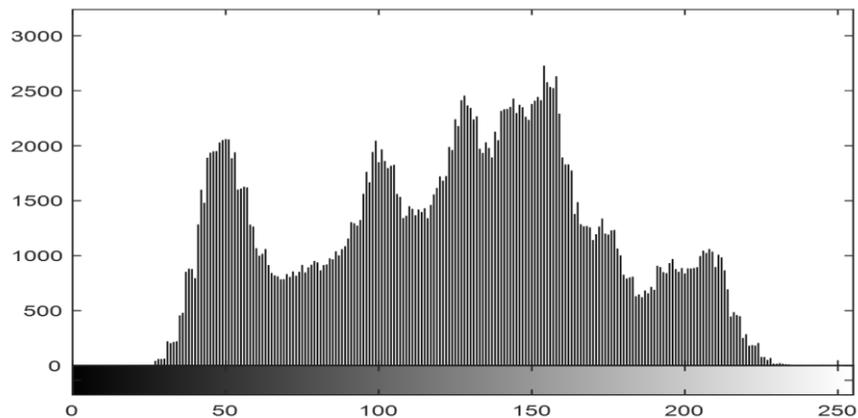

Figure 12 LENA Histogram with 100% ANR Embedding

### 6.2 Embedding Comparison of ANR and Fibonacci

An interpretation of the aforementioned results reveal the resistance of the ANR algorithm to normal statistical attacks. This verifies the original assumptions proposed in this portion of the dissertation research – that splitting a multimedia cover into a non-binary based decomposition and then spreading the covert data over a larger number of non-standard image levels results in a broader dispersion of the covert data with similar decorrelation affects that would be evident in a strictly binary-level image decomposition. This decorrelation affect persist even when the ANR deconstructed image is reconstituted to the original binary form, representable for normal digital processing and display systems.

Additional characteristics of the ANR embedding algorithm are shown when compared to other than LSB embedding techniques. As one example, earlier we defined a Fibonacci redundant number system that is used for several embedding algorithms [15][19][20]. The numerical properties for the decomposition, embedding, and reconstitution image values would suggest that additional embedding capacity exists in 13 levels versus only the 8 standard levels. This is explore in the following cases by comparing several samples of ANR versus Fibonacci embedding.

The first case we show a difference image for the $ANR_{255}$ when 100% of the calculated LSB capacity is used for data embedding. For the *Lena* image, this would be 512x512 or 262,144 data bits. From the full difference images below in Figure 13, it is shown that 100% of the available space only is used in the case of Fibonacci-based embedding while only approximately 60% is actually used in the case of direct $ANR_{255}$ embedding. This shows the ANR embedding capacity is improved by the fact that the information is distributed throughout the bit planes with residual space.

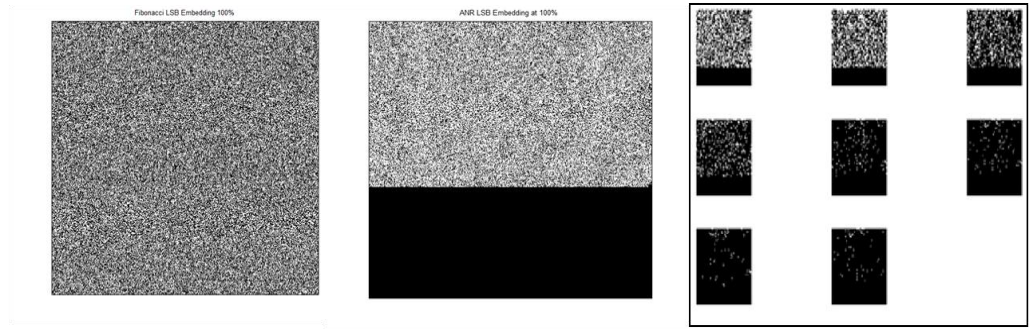

Figure 13 100% Embedding - Fibonacci and 8-bit Difference Decomposition for $ANR_{255}$

## 7. Summary

In this paper, we have reviewed the concepts behind redundant number system data embedding for bitmapped images [15][19][20][22]. We also introduced a new redundant number system embedding algorithm which we refer to as the Adjunctive Numerical Representation or ANR. We then showed, using experimental results, how the ANR system is essentially superior to first and second order statistical steganalysis attacks. Hence, the system has the following advantages:

1. *Provides a novel and unique representation over a redundant number system for any pixel value based on a variety of monotonic number system.*
2. *System is characterized by blind embedding and decoding, eliminating the need for a pre-exchanged dictionary set.*
3. *Higher capacity embedding with minimal impact on visible distortions or statistical anomalies that may indicate the presence of covert data.*

The intent of this research is to introduce the techniques behind the use of alternative number systems data embedding. The capabilities described show the need for more comprehensive development of detection and analysis methods for LSB steganography beyond what is currently in use by cyber defensive systems.

## Authors

**James C. Collins** received a B.S. degree in electrical engineering from Arizona State University, Tempe Arizona in 1987 and a M.S. degree in electrical engineering from Southern Methodist University, Dallas Texas in 1999. He is currently pursuing his Ph.D. degree in electrical engineering, specializing in digital signal processing theory at the University of Texas at San Antonio, Texas. His research interests include embedded systems, network systems security, data hiding, and image processing.

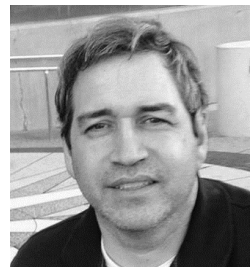

**Sos S. Agaian** is a Peter T. Flawn Professor of Electrical and Computer Engineering at the University of Texas, San Antonio (UTSA), and Professor at the University of Texas Health Science Center, San Antonio. He received a M.S. degree (summa cum laude) in mathematics and mechanics from Yerevan University, Armenia, and a Ph.D. degree in math and physics from the Steklov Institute of Mathematics, Russian Academy of Sciences, as well as a PhD degree in Engineering Sciences from the Institute of the Control System, Russian Academy of Sciences. He has authored more than 500 scientific papers, 7 books, and holds 14 patents. Some of his major research efforts are focused in multimedia processing, imaging systems, information security, artificial intelligence, computer vision, 3D imaging sensors, image fusion, and biomedical and health Informatics.

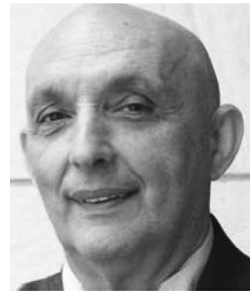